\documentclass[prl,twocolumn,showpacs,preprintnumbers,amsmath,amssymb]{revtex4}
\usepackage{bm}
\usepackage{amsmath}
\usepackage{dcolumn}
\usepackage[dvips]{graphicx}
\begin{document}
\bibliographystyle{apsrev}

\title{Comment on
``Radiative corrections and  parity nonconservation in heavy atoms''}

\author{M. Yu. Kuchiev}
\email[email: ]{ kuchiev@newt.phys.unsw.edu.au} \affiliation{
School of Physics, University of New South Wales, Sydney 2052, Australia}

\author{V. V. Flambaum}
\email[email: ]{ flambaum@newt.phys.unsw.edu.au} \affiliation{
School of Physics, University of New South Wales, Sydney 2052, Australia}

\date{\today}

 \maketitle

A recent paper of Milstein {\it et al}
\cite{milstein_sushkov_terekhov_02},
which is commented below, presents an opportunity
to summarize briefly
interesting results that have been obtained for QED corrections to the
parity nonconservation amplitude (PNC) in atoms.

It has become clear quite recently, that the QED radiate corrections to the
PNC amplitude in atoms are sufficiently large,
of the order of $\sim 1 \%$, and therefore should be
included in an analysis of experimental data.
High accuracy of this data
($0.3 \% $ for the $6s-7s$ PNC amplitude in $^{133}$Cs
\cite{wood_97}) is used as a test of the standard model,
prompting accurate treatment of the radiative corrections.
The most difficult for calculations (and even estimations)
part of the QED corrections presented the self-energy and vertex
corrections to the PNC amplitude.
They also prove to give the largest contribution, as
has become clear only during last several months,
while beforehand  they were assumed to be
negligible small. It should be noted that without
the self-energy corrections the experimental data of Ref. \cite{wood_97}
deviate by more than $2\sigma$ from predictions
of the standard model, for a brief guide  to the modern
literature devoted to the subject  see \cite{kf_02}.
Remarkably, when these corrections are calculated
accurately, their contribution brings the experimental results
of \cite{wood_97} in full agreement with the standard model,
as was first demonstrated in \cite{kf_02} and
confirmed by recent \cite{k_02} and \cite{milstein_sushkov_terekhov_02}.
The above mentioned facts have made last developments in the outlined
research area quite interesting and exciting.

\begin{figure}[t]
\vspace{-2.2cm}
\centering
\includegraphics[height=11cm,width = 2 \textwidth,keepaspectratio=true]{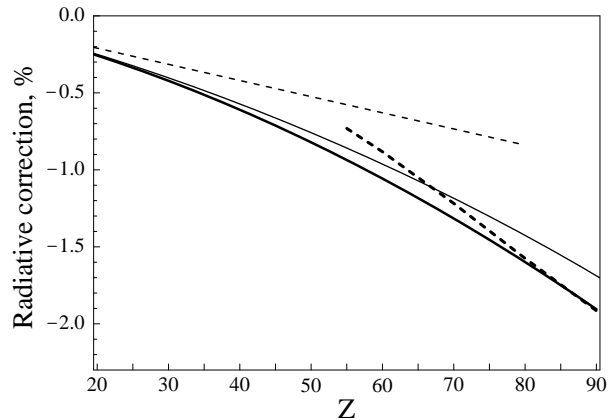}
\vspace{-3cm} \hspace{0pt} \caption{\it { QED self-energy radiative
corrections ($\%$) for the PNC amplitude in atoms
versus the atomic nuclear charge $Z$. The thick short-dashed line,
thick solid line and thin solid line -
results of papers {\rm \cite{kf_02}},
{\rm \cite{k_02} }, and {\rm \cite{milstein_sushkov_terekhov_02}}
respectively. The thin dashed line - the lowest term
of $\alpha Z$ expansion
$ - 1.97 \alpha^2 Z$
{\rm \cite{k_02,milstein_sushkov_terekhov_02}}. }}
\label{fig}
\end{figure}

Fig. \ref{fig} presents a summary of results for the QED self-energy
(plus vertex) radiative corrections to the PNC amplitude in atoms.
It shows these corrections in relative units, as a percent
of the PNC amplitude, versus the charge of the atomic nucleus considered.
The most interesting for recent experimental applications
atoms are Cs, $Z=55$, and  Tl, Pb and Bi
with $Z$ close to $80$. Fig. \ref{fig} shows
a close agreement of results of our Ref.
\cite{kf_02} with the more recent results of
\cite{k_02,milstein_sushkov_terekhov_02}.
The important conclusion that follows from this figure is
that the self-energy correction to the PNC amplitude is negative
and large. For the Cs atom there exists perfect agreement
between all the results
$-0.73(20) \% $ \cite{kf_02}, $-0.9(2) \% $ \cite{k_02}, and $-0.85 \% $
\cite{milstein_sushkov_terekhov_02}.
A slight discrepancy observed for more heavy atoms Tl, Pb, and Bi,
$-1.6 \% $ \cite{kf_02,k_02} and $-1.4 \% $
\cite{milstein_sushkov_terekhov_02},
is, probably, within the errors of the methods used,
as is specified below. As was demonstrated in \cite{kf_02}
and supported by \cite{k_02,milstein_sushkov_terekhov_02},
the found correction brings the experimental data
of Wood {\em et al} \cite{wood_97} within the limits of the standard model.

A reliability of this conclusion is supported by
two different groups which have converged to one and
the same result. This positive development is further magnified
by the fact that previously there has
existed a sharp contradiction between  \cite{kf_02}
and the results of Milstein and Sushkov who insisted that the
self-energy correction is small and positive \cite{milstein_sushkov_02},
casting doubts on possible relation between experimental data of
\cite{wood_97} and the standard model.

Note that the results presented in Fig. \ref{fig}
were obtained in \cite{kf_02,k_02,milstein_sushkov_terekhov_02} using
different methods and approximations. The work \cite{kf_02}
takes into account all orders in the $\alpha Z$-expansion
deriving an identity which relates the radiative corrections
to the PNC matrix elements between the s$_{1/2}$ and p$_{1/2}$
electron wave functions with the radiative corrections to the
finite nuclear size (FNS) energy shifts of
the s$_{1/2}$ and p$_{1/2}$ levels.
The works  \cite{k_02,milstein_sushkov_terekhov_02}
calculated the self-energy corrections to PNC using the $\alpha Z$-expansion.
The results of these latter works can be used to
test validity of the identity found in  Ref. \cite{kf_02}.
Numerical results of \cite{k_02,milstein_sushkov_terekhov_02}
shown in Fig. \ref{fig} indicate that this identity holds,
bringing  the final conclusions of these works
in very close agreement with \cite{kf_02}.
(Note that a possible $\le 20\%$ violation of the identity,
that was proclaimed in
\cite{milstein_sushkov_terekhov_02}, is smaller than 
the error that follows from the parameter
of the perturbative expansion used in
\cite{k_02,milstein_sushkov_terekhov_02},
which is $ \alpha Z=0.4$ for Cs and
$\alpha Z=0.6$ for Tl, Pb, Bi.
The mentioned deviation appeared probably because
the p$_{1/2}$ FNS was extrapolated 
in \cite{milstein_sushkov_terekhov_02}
to small-$Z$ values from the high-$Z$ FNS of the $2p_{1/2}$ state
calculated numerically in different works - see Refs. in
\cite{kf_02,milstein_sushkov_terekhov_02}. This extrapolation 
should incorporate an error governed by 
the parameter mentioned above \cite{difference}.)
The work  \cite{milstein_sushkov_terekhov_02}
calculates also corrections $\sim  \alpha^3 Z^2 \ln mR$ where
$R$ is the nuclear radius verifying (in this order) that
the identity of \cite{kf_02} holds exactly, as it should.

Thus, the results obtained by completely different methods agree.
We believe this signifies a point
after which the discrepancy between
the atomic experimental data of \cite{wood_97}
and the standard model disappears.

This work was supported by the Australian Research Council.

\end{document}